\documentclass[superscriptaddress,column,showpacs,a4paper,
amssymb,amsmath,nobibnotes,aps,prd,longbibliography,
showkeys,
nofootinbib,notitlepage]{revtex4-1}
\usepackage{graphicx}
\usepackage{float}
\usepackage{epsf}
\usepackage{bm}
\usepackage{amsmath}
\usepackage{amsfonts}
\usepackage{amssymb}
\usepackage{epstopdf}
\usepackage{natbib}
\usepackage{color}%
\usepackage{mathrsfs}
\usepackage{bm}
\usepackage{mathtools}
\providecommand{\U}[1]{\protect\rule{.1in}{.1in}}

\newcommand{\be}{\begin{equation}}
\newcommand{\ee}{\end{equation}}

\newcommand{\mincir}{\raise
-3.truept\hbox{\rlap{\hbox{$\sim$}}\raise4.truept\hbox{$<$}\ }}
\newcommand{\magcir}{\raise
-3.truept\hbox{\rlap{\hbox{$\sim$}}\raise4.truept\hbox{$>$}\ }}

\newtheorem{remark}{Remark}[section]
\usepackage{xcolor}
\usepackage{hyperref}
\hypersetup{
    colorlinks=true, 
    linkcolor=blue,
    citecolor=magenta}

\begin{document}

\title{Algebraic representation of General Relativity in harmonic gauge}

\author{Jaume de Haro}
\email{jaime.haro@upc.edu}
\affiliation{Departament de Matem\`atiques, Universitat Polit\`ecnica de Catalunya, Diagonal 647, 08028 Barcelona, Spain}

\begin{abstract}
{
We present a self-contained coordinate-matrix formulation of General Relativity in the harmonic gauge, providing a compact, index-free representation of the gravitational field equations within the algebra of $4\times 4$ real matrices $M_4$. By defining musical isomorphism operators ($\sharp, \flat$), metric adjoints ($\dagger$), and matrix-valued derivative vectors ($\bm{\mathcal{B}}, \bm{\mathcal{C}}, \bm{\mathcal{D}}^\sharp$), the Ricci tensor is recast into an algebraic matrix structure dominated by the Laplace--Beltrami operator $\Box_{\mathrm{LB}} \coloneqq \operatorname{tr}(\mathcal{G}^\sharp \partial^2)$, where $\partial^2 \coloneqq \bm{\partial}\bm{\partial}^T$ represents the matrix Hessian operator of second partial derivatives. Within this setting, Einstein's field equations naturally take the form of a non-linear wave equation for the metric matrix $\mathcal{G}$, giving rise to an index-free matrix representation of the gravitational energy-momentum pseudotensor $\bar{\mathcal{T}}$. We derive the explicit algebraic relation connecting $\bar{\mathcal{T}}$ with the classical Landau--Lifshitz energy-momentum complex $\mathcal{T}_{\mathrm{LL}}$, and express the associated local conservation law in a simple matrix divergence form, $\bm{\partial}^T \left((-g)(\mathcal{T}^{\sharp} + \mathcal{T}_{\mathrm{LL}}^{\sharp})\right) = \bm{0}^T$. Finally,
we illustrate the scope and applicability
 of the matrix formalism through three physical applications: weak-field linearized gravity, the exact derivation of Friedmann cosmological dynamics in harmonic time, and the complete formulation of linear scalar cosmological perturbations in both the Newtonian and harmonic gauges.}
\end{abstract}

\vspace{0.5cm}



\maketitle

{

\section{Introduction}\label{sec:intro}

General Relativity (GR) is traditionally formulated within 
the framework of differential geometry on pseudo-Riemannian manifolds~\cite{MTW1973,Synge1960,Hawking1973,Sachs1977,Carroll2019}, where field equations and geometric tensors are expressed either in index-based component notation or in abstract coordinate-free exterior calculus. While index notation is manifestly covariant, it can frequently obscure the intrinsic linear-algebraic symmetries, matrix traces, and operator structures of the gravitational field equations. This becomes particularly evident when fixing a specific coordinate gauge, such as the harmonic (or wave) gauge, where Einstein's equations simplify into a quasi-linear system of wave equations for the metric components~\cite{Thorne}.

Harmonic coordinates, introduced originally by de Donder~\cite{deDonder1921} and Fock~\cite{Fock1964}, play a decisive role in proving the well-posedness of the Cauchy problem in General Relativity~\cite{ChoquetBruhat2009}, as well as in post-Newtonian expansions~\cite{Will} and gravitational wave generation~\cite{Thorne, Blanchet2014}. In this gauge, coordinate freedom is fixed by requiring each coordinate function to satisfy the scalar wave equation, which effectively isolates the curved-space d'Alembertian (Laplace--Beltrami) operator acting componentwise on the metric, separating it cleanly from quadratic first-derivative terms~\cite{Haro2026,Haro2026a}.

In the spirit of alternative algebraic reformulations of fundamental field equations—such as the Duffin--Kemmer--Petiau matrix formalism for relativistic fields~\cite{Kemmer1939,Duffin1938} or Geometric Algebra formulations of electrodynamics and gravitation~\cite{Hestenes1966,Hestenes2003,Lasenby1998,Doran2003}—the matrix framework presented here does not modify the physical content of General Relativity. Instead, it provides a streamlined, index-free representation specifically optimized for the harmonic gauge, simplifying higher-order derivative structures and facilitating symbolic algebra implementations.

More precisely, our main goal is to construct a rigorous, index-free coordinate-matrix formalism for General Relativity in harmonic coordinates. 
By representing second-order
tensors and connection-related derivative terms onto the space of $4\times 4$ real matrices $M_4$, we establish a compact matrix algebra equipped with metric adjoints ($\dagger$), musical isomorphism operators ($\sharp, \flat$), and matrix trace operations. Within this scheme, the Laplace--Beltrami operator assumes the transparent form $\Box_{\mathrm{LB}} \coloneqq \operatorname{tr}(\mathcal{G}^\sharp \partial^2)$, constructed directly from the column derivative vector $\bm{\partial}$ and the inverse metric matrix $\mathcal{G}^\sharp$.

Using this matrix representation, we demonstrate that Einstein's field equations reduce to a non-linear matrix wave equation, $\Box_{\mathrm{LB}}\mathcal{G} = -16\pi G_{\mathrm{N}} \hat{\mathcal{P}}_-(\mathcal{T} + \bar{\mathcal{T}})$, which naturally defines an index-free matrix gravitational energy-momentum pseudotensor $\bar{\mathcal{T}}$. Furthermore, we derive the exact algebraic connection between $\bar{\mathcal{T}}$ and the classical Landau--Lifshitz energy-momentum complex $\mathcal{T}_{\mathrm{LL}}$~\cite{Landau,Poisson}, casting their interplay and local conservation laws purely in terms of matrix operations. To highlight the analytical power of our framework, we apply it to three concrete physical scenarios: weak-field linearized gravity, the exact derivation of Friedmann cosmological equations in harmonic time, and
a matrix formulation of linear scalar cosmological perturbations
in both the Newtonian and harmonic gauges.

The paper is structured as follows: In Section~\ref{sec:notation}, we lay down the matrix algebra, musical operators, metric adjoints, and matrix-valued derivative vectors. In Section~\ref{sec:field_equations}, we formulate the Ricci matrix in harmonic coordinates, set up the matrix wave equation, establish the link with the Landau--Lifshitz pseudotensor, and apply the framework to linearized gravity and FLRW background cosmology. In Section~\ref{sec:cosmology}, we extend the matrix operators to cosmological scalar perturbations, developing the governing perturbation and conservation equations in both the Newtonian and harmonic gauges. Finally, Section~\ref{sec:conclusions} summarizes our main findings and outlines future applications.

\

Throughout this work, we adopt geometric units where $c=1$, and $G_{\mathrm{N}}$ denotes Newton's gravitational constant.

}

\section{Notation and conventions}\label{sec:notation}

{

Let $(\mathcal{M}, \mathfrak{g})$ be a four-dimensional pseudo-Riemannian manifold representing spacetime. We denote by $T^0_2(\mathcal{M})$ the space of $2$-covariant tensors and by $T^2_0(\mathcal{M})$ the space of $2$-contravariant tensors. Throughout this work, Greek indices $\mu, \nu, \dots \in \{0, 1, 2, 3\}$ label local coordinate components, and Einstein's summation convention over repeated indices is implicitly assumed.

In a local coordinate chart $\{x^\mu\}$, the metric tensor $\mathfrak{g} \in T^0_2(\mathcal{M})$ and its inverse $\mathfrak{g}^{-1} \in T^2_0(\mathcal{M})$ are expressed as
\begin{equation}
    \mathfrak{g} = g_{\mu\nu} dx^{\mu} \otimes dx^{\nu},
\end{equation}
and
\begin{equation}
    \mathfrak{g}^{-1} = g^{\mu\nu} \partial_{\mu} \otimes \partial_{\nu},
\end{equation}
respectively, satisfying $g^{\mu\alpha} g_{\alpha\nu} = \delta^\mu_\nu$.

Two fundamental geometric objects in General Relativity belong to $T^0_2(\mathcal{M})$: the Ricci tensor, $\mathfrak{Ric} \in T^0_2(\mathcal{M})$, and the energy-momentum tensor, $\mathfrak{T} \in T^0_2(\mathcal{M})$, given in coordinate components by
\begin{equation}
    \mathfrak{Ric} = R_{\mu\nu} dx^{\mu} \otimes dx^{\nu},
\end{equation}
and
\begin{equation}
    \mathfrak{T} = T_{\mu\nu} dx^{\mu} \otimes dx^{\nu}.
\end{equation}

For any $2$-covariant tensor $\mathfrak{A} = A_{\mu\nu} dx^{\mu} \otimes dx^{\nu} \in T^0_2(\mathcal{M})$, its metric trace $\operatorname{Tr}(\mathfrak{A})$ is defined by
\begin{equation}
    \operatorname{Tr}(\mathfrak{A}) \coloneqq g^{\mu\nu} A_{\mu\nu}.
\end{equation}
In particular, $\operatorname{Tr}(\mathfrak{g}) = 4$.

Equipped with the metric trace, we introduce the linear trace-reversing operators $\hat{\mathcal{P}}_{\pm} \colon T^0_2(\mathcal{M}) \longrightarrow T^0_2(\mathcal{M})$, defined as
\begin{equation}
    \hat{\mathcal{P}}_{\pm}(\mathfrak{A}) \coloneqq \mathfrak{A} \pm \frac{1}{2}\mathfrak{g} \operatorname{Tr}(\mathfrak{A}).
\end{equation}
The operator $\hat{\mathcal{P}}_-$ satisfies the involutive property $\hat{\mathcal{P}}^2_- = \operatorname{Id}$, where $\operatorname{Id}$ denotes the identity transformation on $T^0_2(\mathcal{M})$. In particular, applying $\hat{\mathcal{P}}_-$ to the Ricci tensor yields the Einstein tensor,
\begin{equation}
    \mathfrak{G} \coloneqq \hat{\mathcal{P}}_-(\mathfrak{Ric}) = \mathfrak{Ric} - \frac{1}{2}\mathfrak{g}\operatorname{Tr}(\mathfrak{Ric}).
\end{equation}

}

{

\subsection{Matrix Representation}\label{subsec:matrix_rep}

Once a local coordinate chart 
$\{x^\mu\}$
has been fixed
 on $\mathcal{M}$, second-order tensors are represented by $4 \times 4$ real matrices in $M_4$. In particular, for the symmetric geometric and physical tensors $\mathfrak{g}$, $\mathfrak{g}^{-1}$, $\mathfrak{Ric}$, and $\mathfrak{T}$, we associate the corresponding symmetric matrices in $M_4$:
\begin{enumerate}
    \item $\mathcal{G} \in M_4$ is the metric matrix with entries $(\mathcal{G})_{\mu\nu} \coloneqq g_{\mu\nu}$, whose determinant $\operatorname{det}(\mathcal{G})$ is denoted by $g$.
    \item $\mathcal{G}^{-1} \in M_4$ is the inverse metric matrix with entries $(\mathcal{G}^{-1})^{\mu\nu} \coloneqq g^{\mu\nu}$.
    \item $\mathcal{R} \in M_4$ is the Ricci matrix with entries $(\mathcal{R})_{\mu\nu} \coloneqq R_{\mu\nu}$.
    \item $\mathcal{T} \in M_4$ is the energy-momentum matrix with entries $(\mathcal{T})_{\mu\nu} \coloneqq T_{\mu\nu}$.
\end{enumerate}

We extend the musical isomorphism operators to $M_4$ as follows:
\begin{enumerate}
    \item \emph{Sharp operator} $\sharp \colon M_4 \longrightarrow M_4$: For any matrix $\mathcal{A} \in M_4$, we define
    \begin{equation}
        \mathcal{A}^{\sharp} \coloneqq \mathcal{G}^{-1}\mathcal{A}\mathcal{G}^{-1}.
    \end{equation}
    In particular, $\mathcal{G}^{\sharp} = \mathcal{G}^{-1}$.

    \item \emph{Flat operator} $\flat \colon M_4 \longrightarrow M_4$: For any matrix $\mathcal{A} \in M_4$, we define
    \begin{equation}
        \mathcal{A}^{\flat} \coloneqq \mathcal{G}\mathcal{A}\mathcal{G}.
    \end{equation}
    In particular, $(\mathcal{G}^{-1})^{\flat} = \mathcal{G}$.
\end{enumerate}

The symmetrization of a matrix $\mathcal{A} \in M_4$ is given by
\begin{equation}
    \operatorname{Sym}(\mathcal{A}) \coloneqq \frac{1}{2}\left(\mathcal{A} + \mathcal{A}^T\right),
\end{equation}
where $\mathcal{A}^T$ denotes the transpose matrix of $\mathcal{A}$.

In analogy with tensor contractions, the metric trace of a matrix $\mathcal{A} \in M_4$ is defined as
\begin{equation}
    \operatorname{Tr}(\mathcal{A}) \coloneqq \operatorname{tr}(\mathcal{G}^{-1}\mathcal{A}),
\end{equation}
where $\operatorname{tr}$ denotes the standard matrix trace on $M_4$.

Next, given two column four-vectors $\bm{v}, \bm{u} \in \mathbb{R}^4$, their metric inner product is defined by
\begin{equation}
    \langle \bm{v}, \bm{u} \rangle_{\mathcal{G}} \coloneqq \bm{v}^T \mathcal{G} \bm{u}.
\end{equation}
Using this inner product, the metric adjoint $\mathcal{A}^{\dagger}$ of a matrix $\mathcal{A} \in M_4$ is defined via
\begin{equation}
    \langle \mathcal{A}\bm{v}, \bm{u} \rangle_{\mathcal{G}} = \langle \bm{v}, \mathcal{A}^{\dagger}\bm{u} \rangle_{\mathcal{G}} 
    \implies 
    \mathcal{A}^{\dagger} \coloneqq \mathcal{G}^{\sharp} \mathcal{A}^T \mathcal{G}.
\end{equation}

Let $\bm{\partial} = (\partial_0, \partial_1, \partial_2, \partial_3)^T$ denote the column vector of partial derivative operators. For any matrix function $\mathcal{A}(x)$, we define the following derivative operations:
\begin{enumerate}
    \item $\bm{\partial}\mathcal{A}$ is the matrix-valued vector whose $\mu$-th component is $\partial_{\mu}\mathcal{A}$, corresponding to entrywise differentiation.
    \item $\bm{\partial}^T\mathcal{A}$ is the matrix-valued row vector resulting from the matrix product of the row vector $\bm{\partial}^T$ and $\mathcal{A}$.
    \item $\bm{\partial\mathcal{A}}$ is the matrix-valued vector whose $\mu$-th component is the matrix with entries
    \begin{equation}
        \left[(\bm{\partial\mathcal{A}})_\mu\right]_{\alpha}^{\beta} \coloneqq \partial_{\alpha}\mathcal{A}^{\beta}_{\mu}.
    \end{equation}
\end{enumerate}

With these definitions in hand, we construct three fundamental matrix-valued derivative vectors that play a central role in Section~\ref{sec:field_equations}:
\begin{equation}
\begin{aligned}
\text{(i)}\quad \bm{\mathcal{B}} &\coloneqq \mathcal{G}^\sharp \bm{\partial}\mathcal{G}, \\[4pt]
\text{(ii)}\quad \bm{\mathcal{C}} &\coloneqq \mathcal{G}^\sharp \bm{\partial\mathcal{G}}, \\[4pt]
\text{(iii)}\quad \bm{\mathcal{D}}^{\sharp} &\coloneqq \bm{\partial\bigl(\sqrt{-g}\,\mathcal{G}^\sharp\bigr)}.
\end{aligned}
\end{equation}

Each component $\mathcal{B}_\mu$ is self-adjoint with respect to the metric adjoint $\dagger$, whereas the components $\mathcal{C}_\mu$ are generally non-self-adjoint. We thus decompose $\bm{\mathcal{C}}$ into its self-adjoint and anti-self-adjoint parts,
\begin{equation}
    \bm{\mathcal{C}}^{(+)} \coloneqq \frac{1}{2}\left(\bm{\mathcal{C}} + \bm{\mathcal{C}}^{\dagger}\right), \qquad 
    \bm{\mathcal{C}}^{(-)} \coloneqq \frac{1}{2}\left(\bm{\mathcal{C}} - \bm{\mathcal{C}}^{\dagger}\right),
\end{equation}
such that $\bm{\mathcal{C}} = \bm{\mathcal{C}}^{(+)} + \bm{\mathcal{C}}^{(-)}$.

Furthermore, given two matrix-valued vectors $\bm{\mathcal{X}} = (\mathcal{X}_\mu)_{\mu=0}^3$ and $\bm{\mathcal{Y}} = (\mathcal{Y}_\mu)_{\mu=0}^3$, we define $\operatorname{tr}(\bm{\mathcal{X}} \otimes \bm{\mathcal{Y}})$ as the matrix whose $(\mu, \nu)$ entry is
\begin{equation}
    \bigl[\operatorname{tr}(\bm{\mathcal{X}} \otimes \bm{\mathcal{Y}})\bigr]^{\mu}_{\nu} \coloneqq \operatorname{tr}(\mathcal{X}_\mu \mathcal{Y}_\nu).
\end{equation}
The matrix $\operatorname{tr}(\bm{\mathcal{X}} \otimes \bm{\mathcal{X}})$ is manifestly symmetric. Its symmetrized generalization for distinct vectors is defined componentwise as
\begin{equation}
    \operatorname{Sym}\bigl[\operatorname{tr}(\bm{\mathcal{X}} \otimes \bm{\mathcal{Y}})\bigr]^{\mu}_{\nu} \coloneqq \frac{1}{2}\left[\operatorname{tr}(\mathcal{X}_\mu \mathcal{Y}_\nu) + \operatorname{tr}(\mathcal{Y}_\mu \mathcal{X}_\nu)\right],
\end{equation}
where equality follows from the cyclic property of the standard matrix trace.

Similarly, we extend the trace-reversing operators $\hat{\mathcal{P}}_{\pm}$ to $M_4$ via
\begin{equation}
    \hat{\mathcal{P}}_{\pm}(\mathcal{A}) \coloneqq \mathcal{A} \pm \frac{1}{2}\mathcal{G}\operatorname{Tr}(\mathcal{A}).
\end{equation}

Finally, defining the coordinate displacement column vector $d\mathbf{x} \coloneqq (dx^0, dx^1, dx^2, dx^3)^T$, the spacetime line element, which physically represents the \textit{infinitesimal proper time},  is given by
\begin{equation}
    ds \coloneqq \sqrt{d\mathbf{x}^T \mathcal{G} d\mathbf{x}}.
\end{equation}
In Minkowski spacetime, where $\mathcal{G} = \mathcal{G}_{\mathrm{Min}} \coloneqq \operatorname{diag}(1, -1, -1, -1)$ \cite{Minkowski1909}, this recovers the flat line element 
\begin{equation}
    ds_{\mathrm{Min}} \coloneqq \sqrt{(dx^0)^2 - (dx^1)^2 - (dx^2)^2 - (dx^3)^2}.
\end{equation}

}

{

\section{Field Equations}\label{sec:field_equations}

Einstein’s field equations \cite{Einstein1916} can be expressed in covariant tensor form as
\begin{equation}
    \mathfrak{Ric} = 8\pi G_{\mathrm{N}} \hat{\mathcal{P}}_-(\mathfrak{T}) \iff \mathfrak{G} = 8\pi G_{\mathrm{N}} \mathfrak{T}.
\end{equation}
These field equations are compatible with the local energy-momentum conservation law
\begin{equation}
    \operatorname{div}_{\mathfrak{g}}(\mathfrak{T}) = 0,
\end{equation}
where $\operatorname{div}_{\mathfrak{g}}$ denotes the metric divergence operator. In local coordinates, this conservation law takes the standard component form
\begin{equation}
    \nabla_{\mu} T^{\mu\nu} = 0,
\end{equation}
where $\nabla$ represents the Levi-Civita connection associated with the metric $\mathfrak{g}$.

\subsection{Ricci Matrix in Harmonic Coordinates}\label{subsec:ricci_harmonic}

Harmonic coordinates (also known as the de Donder or wave gauge) \cite{deDonder1921} are defined by requiring each spacetime coordinate function $x^\mu$ to satisfy the scalar wave equation $\Box_{\mathrm{LB}} x^\mu = 0$, where $\Box_{\mathrm{LB}}$ is the Laplace--Beltrami operator. In coordinate components, this condition is equivalent to the vanishing of the contracted Christoffel symbols,
\begin{equation}
    \Gamma^\mu \coloneqq g^{\alpha\beta} \Gamma^\mu_{\alpha\beta}
      =0\Longrightarrow  \frac{1}{\sqrt{-g}} \partial_\alpha \left(\sqrt{-g} g^{\alpha\mu}\right) = 0
      \Longrightarrow   
g^{\alpha\beta}\left(
\partial_{\alpha}g_{\mu\beta}-\frac{1}{2}\partial_{\mu}g_{\alpha\beta}
\right)=0
      , \quad \text{for } \mu = 0, 1, 2, 3.
\end{equation}

The primary advantage of the harmonic gauge is that it fixes the coordinate gauge freedom by recasting the non-linear system of Einstein's field equations into a system of quasi-linear wave equations. Specifically, the second-order derivative terms of the Ricci tensor $R_{\mu\nu}$ reduce to a pure d'Alembertian operator acting componentwise on the metric, $-\frac{1}{2}g^{\alpha\beta}\partial^2_{\alpha\beta}g_{\mu\nu}$, thereby eliminating all mixed second derivatives.

In harmonic coordinates, the Ricci tensor components take the explicit form \cite{Haro2026,Haro2026a}
\begin{equation}\label{eq:Ricci_components}
\begin{split}
    R_{\mu\nu} = & -\frac{1}{2} g^{\alpha\beta} \partial^2_{\alpha\beta} g_{\mu\nu} - \frac{1}{2} \left(\partial_{\mu}g^{\alpha\beta}\partial_{\alpha}g_{\nu\beta} + \partial_{\nu}g^{\alpha\beta}\partial_{\alpha}g_{\mu\beta}\right) \\
    & - \frac{1}{4} g^{\beta\gamma} g^{\alpha\epsilon} \partial_{\mu}g_{\gamma\alpha} \partial_{\nu}g_{\beta\epsilon} - \frac{1}{2} g^{\beta\gamma} g^{\alpha\epsilon} \partial_{\alpha}g_{\mu\gamma} \left(\partial_{\beta}g_{\nu\epsilon} - \partial_{\epsilon}g_{\nu\beta}\right),
\end{split}
\end{equation}
which we now translate into our coordinate-matrix algebra.

To this end, we express each derivative term on the right-hand side of Eq.~\eqref{eq:Ricci_components} in matrix notation:
\begin{enumerate}
    \item \emph{Second-order derivative term:}
    \begin{equation}
        -\frac{1}{2} g^{\alpha\beta}\partial^2_{\alpha\beta} g_{\mu\nu} = -\frac{1}{2} \left[\Box_{\mathrm{LB}}\mathcal{G}\right]^{\mu}_{\nu},
    \end{equation}
    where the Laplace--Beltrami operator in harmonic coordinates assumes the compact trace structure
    \begin{equation}
        \Box_{\mathrm{LB}} \coloneqq \operatorname{tr}(\mathcal{G}^{\sharp}\partial^2),
    \end{equation}
    with $\partial^2 \coloneqq \bm{\partial}\bm{\partial}^T$ denoting the $4\times 4$ matrix operator of second partial derivatives.

    \item \emph{First derivative term in $\bm{\mathcal{B}}$ and $\bm{\mathcal{C}}^{(+)}$:}
    \begin{equation}
        -\frac{1}{2} \left(\partial_{\mu}g^{\alpha\beta}\partial_{\alpha}g_{\nu\beta} + \partial_{\nu}g^{\alpha\beta}\partial_{\alpha}g_{\mu\beta}\right) = \operatorname{Sym}\left[\operatorname{tr}\left(\mathcal{B}_{\mu} \mathcal{C}_{\nu}^{(+)}\right)\right],
    \end{equation}
    where we have used the identity $\partial_{\mu}g^{\alpha\beta} = -g^{\alpha\sigma}g^{\beta\gamma}\partial_{\mu}g_{\sigma\gamma}$ together with the trace identity $\operatorname{tr}\left(\mathcal{B}_{\mu}\mathcal{C}_{\nu}^{\dagger}\right) = \operatorname{tr}\left(\mathcal{B}_{\mu}\mathcal{C}_{\nu}\right)$.

    \item \emph{Quadratic term in $\bm{\mathcal{B}}$:}
    \begin{equation}
        -\frac{1}{4} g^{\beta\gamma} g^{\alpha\epsilon} \partial_{\mu}g_{\gamma\alpha} \partial_{\nu}g_{\beta\epsilon} = -\frac{1}{4} \operatorname{tr}(\mathcal{B}_{\mu}\mathcal{B}_{\nu}).
    \end{equation}

    \item \emph{Anti-self-adjoint quadratic term in $\bm{\mathcal{C}}^{(-)}$:}
    \begin{equation}
        -\frac{1}{2} g^{\beta\gamma} g^{\alpha\epsilon} \partial_{\alpha}g_{\mu\gamma} \left(\partial_{\beta}g_{\nu\epsilon} - \partial_{\epsilon}g_{\nu\beta}\right) = -\frac{1}{2} \left[\operatorname{tr}\left(\mathcal{C}_{\mu}\mathcal{C}_{\nu}\right) - \operatorname{tr}\left(\mathcal{C}^{\dagger}_{\mu}\mathcal{C}_{\nu}\right)\right] = -\operatorname{tr}\left(\mathcal{C}_{\mu}^{(-)} \mathcal{C}_{\nu}^{(-)}\right),
    \end{equation}
    where we have exploited the property $\operatorname{tr}\left(\mathcal{C}_{\mu}\mathcal{C}_{\nu}\right) = \operatorname{tr}\left(\mathcal{C}_{\mu}^{\dagger}\mathcal{C}_{\nu}^{\dagger}\right)$.
\end{enumerate}

Combining these individual identities, the Ricci matrix $\mathcal{R}$ in harmonic coordinates takes the concise matrix representation
\begin{equation}
    \mathcal{R} = -\frac{1}{2}\Box_{\mathrm{LB}}\mathcal{G} - \frac{1}{4}\operatorname{tr}\left(\bm{\mathcal{B}}\otimes\bm{\mathcal{B}} + 4\bm{\mathcal{C}}^{(-)}\otimes\bm{\mathcal{C}}^{(-)}\right) + \operatorname{Sym}\left[\operatorname{tr}\left(\bm{\mathcal{B}}\otimes\bm{\mathcal{C}}^{(+)}\right)\right].
\end{equation}

Next, we define the gravitational energy-momentum pseudotensor matrix $\bar{\mathcal{T}}$, which contains exclusively first-order derivative terms, as
\begin{equation}
    \bar{\mathcal{T}} \coloneqq -\frac{1}{8\pi G_{\mathrm{N}}} \hat{\mathcal{P}}_-\left(\mathcal{R} + \frac{1}{2}\Box_{\mathrm{LB}}\mathcal{G}\right).
\end{equation}

With this definition, Einstein’s field equations take the form of a non-linear matrix wave equation for the metric matrix $\mathcal{G}$ \cite{Thorne}:
\begin{equation}\label{eq:matrix_wave_equation}
    \Box_{\mathrm{LB}}\mathcal{G} = -16\pi G_{\mathrm{N}} \hat{\mathcal{P}}_-(\mathcal{T} + \bar{\mathcal{T}}) \iff \hat{\mathcal{P}}_-(\Box_{\mathrm{LB}}\mathcal{G}) = -16\pi G_{\mathrm{N}}(\mathcal{T} + \bar{\mathcal{T}}).
\end{equation}

}

{

\subsubsection{The Linear Approximation}\label{subsec:linear_approx}

We consider a perfect fluid source described by the energy-momentum matrix
\begin{equation}\label{eq:stress_matrix}
    \mathcal{T} = (\rho+p)\left(\bm{v}\bm{v}^T\right)^{\flat} - p\mathcal{G} 
    \implies 
    \operatorname{Tr}(\mathcal{T}) = \rho - 3p,
\end{equation}
where $\bm{v}$ is the fluid four-velocity satisfying the metric normalization $\langle \bm{v}, \bm{v}\rangle_{\mathcal{G}} = 1$, and the scalar quantities $\rho$ and $p$ represent the energy density and isotropic pressure measured in the comoving frame, respectively.

In the weak-field approximation, we describe the metric as a small perturbation $\mathcal{H}$ of the Minkowski background
 $\mathcal{G}_{\mathrm{Min}}$:
\begin{equation}\label{eq:metric_perturbation}
    \mathcal{G} \coloneqq \mathcal{G}_{\mathrm{Min}} + \mathcal{H},
\end{equation}
where $\mathcal{G}_{\mathrm{Min}} \coloneqq \operatorname{diag}(1, -1, -1, -1)$.

Substituting Eq.~\eqref{eq:metric_perturbation} into the non-linear matrix wave equation \eqref{eq:matrix_wave_equation} and expanding to first order in $\mathcal{H}$, we obtain the linearized wave equation
\begin{equation}
    \Box \mathcal{H} = -16\pi G_{\mathrm{N}} \left(\mathcal{T}_{\mathrm{Min}} - \frac{1}{2}(\rho-3p)\mathcal{G}_{\mathrm{Min}}\right),
\end{equation}
where $\Box \coloneqq \operatorname{tr}(\partial^2)$ denotes the flat d'Alembertian operator, and $\mathcal{T}_{\mathrm{Min}}$ is the energy-momentum matrix evaluated in Minkowski spacetime,
\begin{equation}
    \mathcal{T}_{\mathrm{Min}} = (\rho+p)\mathcal{G}_{\mathrm{Min}}(\bm{v}_{\mathrm{Min}}\bm{v}_{\mathrm{Min}}^T)\mathcal{G}_{\mathrm{Min}} - p\mathcal{G}_{\mathrm{Min}},
\end{equation}
with the background velocity satisfying $\langle \bm{v}_{\mathrm{Min}}, \bm{v}_{\mathrm{Min}}\rangle_{\mathcal{G}_{\mathrm{Min}}} \coloneqq \bm{v}_{\mathrm{Min}}^T \mathcal{G}_{\mathrm{Min}}\bm{v}_{\mathrm{Min}} = 1$.

Explicitly, the wave equation takes the form
\begin{equation}
    \Box \mathcal{H} = -16\pi G_{\mathrm{N}} \mathcal{G}_{\mathrm{Min}} \left((\rho+p)(\bm{v}_{\mathrm{Min}}\bm{v}_{\mathrm{Min}}^T) - \frac{1}{2}(\rho-p)\mathcal{G}_{\mathrm{Min}}^{-1}\right)\mathcal{G}_{\mathrm{Min}}.
\end{equation}
Exploiting the property $\mathcal{G}_{\mathrm{Min}} = \mathcal{G}_{\mathrm{Min}}^{-1}$ and defining the transformed metric perturbation matrix $\bar{\mathcal{H}} \coloneqq \mathcal{G}_{\mathrm{Min}}\mathcal{H}\mathcal{G}_{\mathrm{Min}}$, the linearized field equations simplify to
\begin{equation}
    \Box \bar{\mathcal{H}} = -16\pi G_{\mathrm{N}} \left((\rho+p)(\bm{v}_{\mathrm{Min}}\bm{v}_{\mathrm{Min}}^T) - \frac{1}{2}(\rho-p)\mathcal{G}_{\mathrm{Min}}\right).
\end{equation}

By parametrizing the metric perturbation matrix $\mathcal{H}$ in terms of scalar, vector, and matrix gravitational potentials as
\begin{equation}
    \mathcal{H} = 
    \left(\begin{array}{c|c}
2\Phi_1+4\mbox{tr}(\mathcal{U}_{(3)}) & -4\mathbf{N}_{(3)}^T \\ \hline
        -4\mathbf{N}_{(3)} &  2\Phi_2\operatorname{Id}_{(3)} + 4\mathcal{U}_{(3)}
    \end{array}\right),
\end{equation}
where $\Phi_1, \Phi_2$ are scalar potentials, $\mathbf{N}_{(3)}$ is a three-dimensional column vector potential, and $\mathcal{U}_{(3)}$ is a $3\times 3$ matrix potential, the linearized wave equations decouple into the system
\begin{equation}\label{eq_lin}
\begin{cases}
    \Box \Phi_1 = -4\pi G_{\mathrm{N}}(\rho + 3p), 
    \\[10pt]
    \Box \Phi_2 = -4\pi G_{\mathrm{N}}(\rho - p), \\[6pt]
    \Box \mathbf{N}_{(3)} = -\dfrac{4\pi G_{\mathrm{N}}(\rho+p)}{1 - |\bm{v}_{(3)}|^2} \bm{v}_{(3)}, \\[10pt]
    \Box \mathcal{U}_{(3)} = -\dfrac{4\pi G_{\mathrm{N}}(\rho+p)}{1 - |\bm{v}_{(3)}|^2} \bm{v}_{(3)}\bm{v}_{(3)}^T,
\end{cases}
\end{equation}
where $\bm{v}_{(3)} \coloneqq \left(\frac{dx}{dt}, \frac{dy}{dt}, \frac{dz}{dt}\right)^T$ represents the spatial three-velocity vector.

{ 
\begin{remark}
    The linearized field equations can be written as follows
    \begin{equation}
\Box\left(\mathcal{G}_{\rm Min}\mathcal{H}-\frac{1}{2}\mbox{tr}(\mathcal{G}_{\rm Min}\mathcal{H})\mbox{Id}\right)=-16\pi G_{\rm N}\mathcal{G}_{\rm Min}\mathcal{T}_{\rm Min}.
    \end{equation}

Using the linearized harmonic condition
\begin{equation}
    \bm{\partial}^T\left(
    \mathcal{G}_{\rm Min}\mathcal{H}-\frac{1}{2}\mbox{tr}(\mathcal{G}_{\rm Min}\mathcal{H})\mbox{Id}_{(4)}
    \right)=\bm{0}^T,
\end{equation}
where using that 
$\mbox{tr}\left(\mathcal{G}_{\rm Min}\mathcal{H}\right)=2(\Phi_1-3\Phi_2)$, can be written as
{
\begin{equation}
    \left\{\begin{array}{ccc}
     \dot{\Phi}_1 +3\dot{\Phi}_2+4\frac{d}{dt}\left(\mbox{Tr}(\mathcal{U}_{(3)})\right) +4\bm{\partial}_{(3)}^T \mathbf{N}_{(3)}  &  =&0, \\
4\dot{\mathbf{N}}_{(3)}^T+  \bm{\partial}_{(3)}^T(\Phi_1-3\Phi_2)+4\bm{\partial}_{(3)}^T\mathcal{U}_{(3)} +2\bm{\partial}_{(3)}^T(\Phi_2\mbox{Id}_{(3)})  & = & \bm{0}_{(3)}^T, 
    \end{array}
    \right.
\end{equation}}
one gets the conservation law
\begin{equation}
\bm{\partial}^T\left(\mathcal{G}_{\rm Min}\mathcal{T}_{\rm Min}\right)=\bm{0}^T.
\end{equation}

\end{remark}

Having established the field equations, the motion of the fluid source is governed by the conservation law in Minkowski spacetime,
\begin{equation}
    \bm{\partial}^T(\mathcal{G}_{\mathrm{Min}}\mathcal{T}_{\mathrm{Min}}\mathcal{G}_{\mathrm{Min}}) = \bm{0}^T 
    \implies 
    \bm{\partial}^T\left((\rho+p)(\bm{v}_{\mathrm{Min}}\bm{v}_{\mathrm{Min}}^T) - p\mathcal{G}_{\mathrm{Min}}\right) = \bm{0}^T,
\end{equation}
supplemented by a barotropic equation of state $p = p(\rho)$. This yields the relativistic continuity and Euler equations \cite{Comer}:
\begin{equation}
    \frac{d\rho}{ds_{\mathrm{Min}}} = -(\rho+p) \bm{\partial}^T \bm{v}_{\mathrm{Min}}, \qquad 
    (\rho+p)\frac{d\bm{v}_{\mathrm{Min}}}{ds_{\mathrm{Min}}} = \bm{\partial}p - \frac{dp}{ds_{\mathrm{Min}}} \bm{v}_{\mathrm{Min}},
\end{equation}
where differentiation is performed with respect to the Minkowski proper time $s_{\mathrm{Min}}$ \cite{Minkowski1909}.

In the static fluid limit, where $\bm{v}_{\mathrm{Min}} = (1, 0, 0, 0)^T$ and $\mathcal{T}_{\mathrm{Min}} = \operatorname{diag}(\rho, p, p, p)$, the d'Alembertian operator reduces to the spatial Laplacian $\Delta$, yielding the static field equation
\begin{equation}
    \Delta \mathcal{H} = 8\pi G_{\mathrm{N}} \operatorname{diag}(\rho+3p, \rho-p, \rho-p, \rho-p).
\end{equation}
Furthermore, energy-momentum conservation in the static regime implies
\begin{equation}
    \bm{\partial}p = \bm{0} \implies p = \text{constant}.
\end{equation}

}

{

\subsubsection{Friedmann Equations in Harmonic Coordinates}\label{subsec:friedmann_harmonic}

In harmonic coordinates, the spatially flat Friedmann--Lemaître--Robertson--Walker (FLRW) metric matrix is expressed as
\begin{equation}\label{eq:FLRW_harmonic_metric}
    \mathcal{G}_{\mathrm{FLRW}}^{\mathrm{har}} = \operatorname{diag}\left(a^6(t), -a^2(t), -a^2(t), -a^2(t)\right),
\end{equation}
where $t$ denotes harmonic time and $a(t)$ is the cosmological scale factor.

For this metric, the Ricci matrix assumes the diagonal form
\begin{equation}
    \mathcal{R} = \operatorname{diag}\left(-3\left(\dot{H}_{\mathrm{har}} - 2H^2_{\mathrm{har}}\right), \frac{\dot{H}_{\mathrm{har}}}{a^4}, \frac{\dot{H}_{\mathrm{har}}}{a^4}, \frac{\dot{H}_{\mathrm{har}}}{a^4}\right),
\end{equation}
where $H_{\mathrm{har}} \coloneqq \dot{a}/a$ represents the Hubble expansion rate in harmonic coordinates, and dots denote differentiation with respect to harmonic time $t$.

The perfect fluid energy-momentum matrix $\mathcal{T}$ takes the diagonal representation
\begin{equation}
    \mathcal{T} = \operatorname{diag}\left(a^6\rho, a^2p, a^2p, a^2p\right).
\end{equation}

Substituting these expressions into the non-linear matrix wave equation \eqref{eq:matrix_wave_equation}, the spatial diagonal components $(k,k)$ for $k \in \{1, 2, 3\}$ yield
\begin{equation}
    \dot{H}_{\mathrm{har}} = 4\pi G_{\mathrm{N}} a^6 (\rho - p),
\end{equation}
while the temporal component $(0,0)$ gives the second Friedmann equation in harmonic coordinates,
\begin{equation}
    -3\dot{H}_{\mathrm{har}} + 6H_{\mathrm{har}}^2 = 4\pi G_{\mathrm{N}} a^6 (\rho + 3p).
\end{equation}

Combining these two relations straightforwardly yields the first Friedmann equation in the harmonic gauge:
\begin{equation}\label{eq:first_friedmann_harmonic}
    H^2_{\mathrm{har}} = \frac{8\pi G_{\mathrm{N}}}{3} a^{6} \rho.
\end{equation}

\begin{remark}
Expressed in terms of standard cosmic time $t_{\mathrm{c}}$, the Friedmann equations recover their familiar textbook form:
\begin{equation}
    H^2 = \frac{8\pi G_{\mathrm{N}}}{3} \rho, \qquad \frac{a''}{a} = -\frac{4\pi G_{\mathrm{N}}}{3}(\rho + 3p),
\end{equation}
where primes denote differentiation with respect to cosmic time $t_{\mathrm{c}}$. Utilizing the differential time transformation
\begin{equation}
    dt_{\mathrm{c}} = a^3(t) \, dt,
\end{equation}
these cosmic-time field equations map directly into Eqs.~\eqref{eq:FLRW_harmonic_metric}--\eqref{eq:first_friedmann_harmonic}, validating the consistency of our coordinate-matrix formalism.
\end{remark}

}

{

\subsection{The Landau--Lifshitz Pseudotensor in Matrix Notation}\label{subsec:landau_lifshitz}

Energy-momentum pseudotensors provide a local formulation for gravitational energy-momentum conservation in curved spacetime, a subject extensively treated in foundational literature \cite{Landau, MTW1973, Wald1984}. Introduced originally by Einstein within the context of \textit{unimodular relativity} \cite{Einstein1916}, pseudotensors remain a fundamental tool for evaluating local gravitational energy balance. In this subsection, we formulate the classical Landau--Lifshitz pseudotensor \cite{Landau} within our coordinate-matrix algebra.

First, we emphasize that the pseudotensor matrix $\bar{\mathcal{T}}$ defined in Eq.~\eqref{eq:matrix_wave_equation} should not be confused with the matrix associated with the classical Landau--Lifshitz energy-momentum complex \cite{Landau}, denoted by $\mathfrak{T}_{\mathrm{LL}}$. The latter provides a distinct, coordinate-dependent expression for the total energy-momentum conservation law,
\begin{equation}\label{eq:LL_conservation}
    \operatorname{div}\left[(-g)\left(\mathfrak{T} + \mathfrak{T}_{\mathrm{LL}}\right)\right] = 0 
    \iff 
    \partial_{\mu}\left((-g)(T^{\mu\nu} + T^{\mu\nu}_{\mathrm{LL}})\right) = 0,
\end{equation}
where $g \coloneqq \operatorname{det}(\mathcal{G})$. In harmonic coordinates, the components of the Landau--Lifshitz pseudotensor assume the explicit form \cite{Landau}
\begin{equation}
\begin{aligned}
(-g)T^{\mu\nu}_{\mathrm{LL}} \coloneqq \frac{1}{16\pi G_{\mathrm{N}}} \Bigg( & \frac{1}{2} {g}^{\mu\nu} {g}_{\lambda\alpha} \partial_\rho \mathfrak{g}^{\lambda\beta} \partial_\beta \mathfrak{g}^{\alpha\rho} - ({g}^{\mu\lambda} \partial_\rho \mathfrak{g}^{\beta\nu} + {g}^{\lambda\nu} \partial_\rho \mathfrak{g}^{\mu\beta}) g_{\alpha\beta} \partial_\lambda\mathfrak{g}^{\alpha\rho} \\
& + {g}_{\lambda\alpha} {g}^{\beta\rho} \partial_\beta \mathfrak{g}^{\mu\lambda} \partial_\rho \mathfrak{g}^{\alpha\nu} + \frac{1}{8} \left( 2 {g}^{\mu\lambda} {g}^{\alpha\nu} - {g}^{\mu\nu} {g}^{\lambda\alpha} \right) \left( 2 {g}_{\beta\rho} {g}_{\sigma\tau} - {g}_{\rho\sigma} {g}_{\beta\tau} \right) \partial_\lambda \mathfrak{g}^{\beta\tau} \partial_\alpha \mathfrak{g}^{\rho\sigma} \Bigg),
\end{aligned}
\end{equation}
where $\mathfrak{g}^{\alpha\beta} \coloneqq \sqrt{-g}\,g^{\alpha\beta}$ denote the components of the "gothic inverse metric", which satisfy the harmonic gauge condition $\partial_{\alpha}\mathfrak{g}^{\alpha\beta} = 0$.

In tensor notation, introducing the Landau--Lifshitz superpotential $H^{\mu\alpha\nu\beta} \coloneqq \mathfrak{g}^{\mu\nu}\mathfrak{g}^{\alpha\beta} - \mathfrak{g}^{\mu\beta}\mathfrak{g}^{\alpha\nu}$ \cite{Poisson}, Einstein's field equations \eqref{eq:matrix_wave_equation} can be rewritten as
\begin{equation}
    \partial_{\alpha\beta}^2 H^{\mu\alpha\nu\beta} = 16\pi G_{\mathrm{N}} (-g)\left(T^{\mu\nu} + T_{\mathrm{LL}}^{\mu\nu}\right).
\end{equation}
Due to the anti-symmetry properties of the superpotential, $H^{\mu\alpha\nu\beta}$ satisfies the identity
\begin{equation}
    \partial_{\alpha\beta\mu}^3 H^{\mu\alpha\nu\beta} = 0,
\end{equation}
which guarantees the local conservation law in Eq.~\eqref{eq:LL_conservation}.

In terms of the Landau--Lifshitz matrix $\mathcal{T}_{\mathrm{LL}}$, with entries $(\mathcal{T}_{\mathrm{LL}})_{\mu\nu} \coloneqq (T_{\mathrm{LL}})_{\mu\nu}$, the field equations take the compact matrix wave form
\begin{equation}\label{eq:LL_matrix_field_eq}
    \sqrt{-g}\left(\Box_{\mathrm{LB}}\left(\sqrt{-g}{\mathcal{G}}^{\sharp}\right)\right) - \operatorname{tr}\left(\bm{\mathcal{D}}^{\sharp} \otimes \bm{\mathcal{D}}^{\sharp}\right) = 16\pi G_{\mathrm{N}}(-g)\left(\mathcal{T}^{\sharp} + \mathcal{T}_{\mathrm{LL}}^{\sharp}\right),
\end{equation}
and the associated local energy-momentum conservation law assumes the simple matrix divergence form
\begin{equation}
    \bm{\partial}^T \left((-g)\left(\mathcal{T}^{\sharp} + \mathcal{T}_{\mathrm{LL}}^{\sharp}\right)\right) = \bm{0}^T.
\end{equation}

Finally, we establish the explicit algebraic connection between the pseudotensor matrix $\bar{\mathcal{T}}$ and the Landau--Lifshitz matrix $\mathcal{T}_{\mathrm{LL}}$. Applying the flat operator $\flat$ to Eq.~\eqref{eq:LL_matrix_field_eq} and combining it with Eq.~\eqref{eq:matrix_wave_equation}, we arrive at the exact relation
\begin{equation}
    \mathcal{{T}}_{\mathrm{LL}} = \bar{\mathcal{T}} + \frac{1}{16\pi G_{\mathrm{N}}} \left[ \hat{\mathcal{P}}_-\left(\Box_{\mathrm{LB}}\mathcal{G}\right) + \frac{1}{\sqrt{-g}}\left(\Box_{\mathrm{LB}}\left(\sqrt{-g}{\mathcal{G}}^{\sharp}\right)\right)^{\flat} - \frac{1}{(-g)}\operatorname{tr}\left(\bm{\mathcal{D}}^{\sharp} \otimes \bm{\mathcal{D}}^{\sharp}\right)^{\flat} \right].
\end{equation}

}

{

\section{Cosmological Perturbations}\label{sec:cosmology}

Cosmological perturbations are essential for understanding the formation of large-scale structure in the Universe and the anisotropies of the cosmic microwave background (CMB). In this section, we apply the coordinate-matrix framework to scalar cosmological perturbations, expressing the perturbed field equations, matter conservation laws, and gauge conditions in a unified algebraic form.


To facilitate the matrix formulation of differential operators on curved cosmological backgrounds, we first introduce the metric-adapted gradient of a scalar function $f$ and the metric divergence of a four-vector $\bm{v}$:
\begin{equation}
    \operatorname{grad}_{\mathcal{G}}(f) \coloneqq \mathcal{G}^\sharp \bm{\partial}f, \qquad
    \operatorname{div}_{\mathcal{G}}(\bm{v}) \coloneqq \frac{1}{\sqrt{-g}} \bm{\partial}^T \left(\sqrt{-g}\bm{v}\right).
\end{equation}

Using these intrinsic matrix operators, we examine scalar cosmological perturbations in two primary coordinate gauges: the Newtonian (longitudinal) gauge and the harmonic gauge.

\subsection{Newtonian Gauge}\label{subsec:newtonian_gauge}

{
Throughout this subsection, following the results obtained in \cite{Elizalde}, we adopt cosmic-time coordinates $(t_{\mathrm{c}}, \mathbf{r})$, where $t_{\mathrm{c}}$ represents standard cosmic time and $\mathbf{r} = (x, y, z)^{\top}$ denotes spatial Cartesian coordinates.

}

In the Newtonian gauge, scalar metric perturbations are parametrized by a single gravitational potential $\Phi_{\mathrm{N}}$, yielding the diagonal perturbed metric matrix
\begin{equation}
    \mathcal{G} \coloneqq \operatorname{diag}\left(\frac{a^2(t_c)}{a_{\mathrm{N}}^2}, -a_{\mathrm{N}}^2, -a_{\mathrm{N}}^2, -a_{\mathrm{N}}^2\right)\cong \mbox{diag}(1+2\Phi_{\mathrm{N}},a^2(t_c)(1-2\Phi_{\mathrm{N}}),a^2(t_c)(1-2\Phi_{\mathrm{N}}),a^2(t_c)(1-2\Phi_{\mathrm{N}})),
\end{equation}
where $a_{\mathrm{N}} \coloneqq a(t_{\mathrm{c}})\sqrt{1 - 2\Phi_{\mathrm{N}}}$ plays the role of a local \emph{perturbed scale factor}.

To capture the local expansion rate of the perturbed cosmological background, we define the \emph{Hubble four-vector} as
\begin{equation}
    \mathbf{H} \coloneqq \frac{1}{a_{\mathrm{N}}} \operatorname{grad}_{\mathcal{G}}(a_{\mathrm{N}}).
\end{equation}
Taking the metric divergence of the Hubble four-vector and evaluating the fluid energy-momentum contribution, we construct the \emph{perturbed Friedmann equation} \cite{Elizalde}:
\begin{equation}\label{eq:perturbed_friedmann_div}
    \operatorname{div}_{\mathcal{G}}(\mathbf{H}) = \frac{8\pi G_{\mathrm{N}}}{3} \bm{v}^T \hat{\mathcal{P}}_+(\mathcal{T}) \bm{v} = 4\pi G_{\mathrm{N}}(\rho - p),
\end{equation}
where the energy-momentum matrix $\mathcal{T}$ is given by Eq.~\eqref{eq:stress_matrix} and $\bm{v}$ represents the fluid four-velocity vector,
\begin{equation}
    \bm{v} \coloneqq \left(\frac{dt_{\mathrm{c}}}{ds}, \frac{d\mathbf{r}}{ds}\right)^T.
\end{equation}

The dynamics of the matter fluid are dictated by the covariant conservation law of the energy-momentum tensor. In our matrix notation, this yields the relativistic continuity equation,
\begin{equation}\label{eq:matrix_continuity}
    \operatorname{div}_{\mathcal{G}}\left((\rho + p)\bm{v}\right) = \bm{v}^T \bm{\partial}p,
\end{equation}
and the relativistic Euler equation governing momentum transport,
\begin{equation}\label{eq:matrix_euler}
    (\rho + p)\frac{D\bm{v}}{ds} = \mathcal{G}^\sharp \bm{\partial}p - \left(\bm{v}^T \mathcal{G}^\sharp \bm{\partial}p\right)\bm{v} \eqcolon \operatorname{grad}_{\bm{v}}^{\perp}(p),
\end{equation}
where $D\bm{v}/ds \coloneqq \nabla_{\bm{v}}\bm{v}$ denotes the directional covariant derivative along the fluid flow lines, and $\operatorname{grad}_{\bm{v}}^{\perp}(p) \coloneqq \operatorname{grad}_{\mathcal{G}}(p) - (\bm{v}^T\operatorname{grad}_{\mathcal{G}}(p))\bm{v}$ represents the orthogonal projection of the pressure gradient onto the subspace orthogonal to $\bm{v}$.

By introducing the logarithmic local scale factor variable $\phi_{\mathrm{N}} \coloneqq \ln a_{\mathrm{N}}$, the governing field and fluid equations \eqref{eq:perturbed_friedmann_div}--\eqref{eq:matrix_euler} assume the remarkably compact form
\begin{equation}
\Box_{\mathrm{LB}}\phi_{\mathrm{N}} = 4\pi G_{\mathrm{N}}(\rho - p), \qquad  
    \operatorname{div}_{\mathcal{G}}\left((\rho + p)\bm{v}\right) = \bm{v}^T \bm{\partial}p, \qquad 
    (\rho + p)\frac{D\bm{v}}{ds} = \operatorname{grad}_{\bm{v}}^{\perp}(p).
\end{equation}

Expanding these expressions to first order in the perturbation variables,
\begin{equation}
    a_{\mathrm{N}} \approx a(t_{\mathrm{c}})(1 - \Phi_{\mathrm{N}}), \quad 
    \rho \approx \bar{\rho}(t_{\mathrm{c}}) + \delta\rho, \quad 
    p \approx \bar{p}(t_{\mathrm{c}}) + \delta p,
\end{equation}
and linearizing the four-velocity components as
\begin{equation}
    \frac{dt_{\mathrm{c}}}{ds} \approx 1 - 2\Phi_{\mathrm{N}}, \qquad  
    \frac{d\mathbf{r}^T}{ds} \approx \bm{v}_{(3)} \coloneqq \frac{d\mathbf{r}^T}{dt_{\mathrm{c}}},
\end{equation}
we precisely recover the standard linear scalar cosmological perturbation equations of General Relativity \cite{Mukhanov}:
{\begin{eqnarray}\left\{\begin{array}{ccc}
    \Delta\Phi_{\rm N} - 3H\big(\partial_{t_c}\Phi_{\rm N} + H\Phi_{\rm N}\big) &=& 4\pi G_{\rm N}\,\delta\rho,\\
4\pi G_{\rm N}\, a^2 (\bar{\rho} + \bar{p})\, {\bm {v}}_{(3),\parallel} &=& -\bm{\partial}_{(3)}(\partial_{t_c} \Phi_{\rm N} + H\Phi_{\rm N}),\\
    \partial_{t_c^2}\Phi_{\rm N} + 4H\partial_{t_c}\Phi_{\rm N} + (2{H}'+3H^2)\Phi_{\rm N} &=& 4\pi G_{\rm N}\,\delta p,\end{array}\right.
\end{eqnarray}
where ${\bm {v}}_{(3),\parallel}
$ is curl-free component of the spatial velocity.

}

\subsection{Harmonic Gauge}\label{subsec:harmonic_cosmo_gauge}

We now turn to the harmonic gauge, which is especially well-suited for analyzing wave propagation and gravitational radiation on cosmological backgrounds due to its explicit hyperbolic structure. In this subsection, we adopt harmonic spacetime coordinates $(t, \mathbf{r})$, where $t$ represents harmonic time and $\mathbf{r}:=(x,y,z)$ are the spatial harmonic coordinates.

In the harmonic gauge, the full metric matrix is expressed as a perturbation $\mathcal{H}$ propagating on the unperturbed harmonic FLRW background matrix $\mathcal{G}_{\mathrm{FLRW}}^{\mathrm{har}}$ defined in Eq.~\eqref{eq:FLRW_harmonic_metric}:
\begin{equation}
    \mathcal{G} \coloneqq \mathcal{G}_{\mathrm{FLRW}}^{\mathrm{har}} + \mathcal{H}.
\end{equation}

{
We decompose the symmetric perturbation matrix $\mathcal{H}$ into a block structure in terms of a scalar potential $\Psi_1$, a three-dimensional spatial vector potential $\mathbf{N}_{\mathrm{har},(3)}$, and a symmetric $3\times 3$ spatial matrix potential $\mathcal{V}_{\mathrm{har},(3)}$ whose diagonal elements vanish identically:
\begin{equation}
    \mathcal{H} = \left(\begin{array}{c|c}
        2a^6\Psi_1 & -4a^6\mathbf{N}_{\mathrm{har},(3)}^T \\[4pt] \hline \\[-8pt]
-4a^6\mathbf{N}_{\mathrm{har},(3)} & \mathcal{H}_{(3)}\odot\operatorname{Id}_{(3)} 
+ 4a^6 \mathcal{V}_{\mathrm{har},(3)}
    \end{array}\right).
\end{equation}
where $\mathcal{H}_{(3)}$ denotes the $3\times 3$ spatial block of $\mathcal{H}$
and $\odot$ represents the Hadamard (entrywise) matrix product.

In addition, it is convenient to introduce an auxiliary spatial trace scalar potential $\Psi_2$, defined by
\begin{equation}
    \Psi_2 \coloneqq \frac{1}{6a^2}\operatorname{tr}(\mathcal{H}_{(3)}). 
\end{equation}

Evaluating the linearized matrix field equations \eqref{eq:matrix_wave_equation} on the harmonic FLRW background, the perturbed field equations decouple into a structured system for the scalar, vector, and matrix potentials \cite{Haro2026}:
\begin{equation}\label{eq:harmonic_perturbation_system}
\begin{cases}
    -3\Box_{\mathrm{har}}(a^6\Psi_1) + 36 H_{\mathrm{har}}^2\Psi_1 - 24H_{\mathrm{har}}\dot{\Psi}_1 = 12\pi G_{\mathrm{N}} a^6(\delta \rho + 3\delta p), \\[8pt]
    -3a^4\Box_{\mathrm{har}}(a^2\Psi_2) + 6\dot{H}_{\mathrm{har}}(\Psi_1 + \Psi_2) - 12H^2_{\mathrm{har}}(4\Psi_1 + \Psi_2) - 4H_{\mathrm{har}}\dot{\Psi}_1 = 12 \pi G_{\mathrm{N}} a^6(\delta\rho - \delta p), 
    \\[8pt]
    -3\Box_{\mathrm{har}} \left(a^6 \mathbf{N}_{\mathrm{har,(3)}}\right) + 12 H_{\mathrm{har}} \left(\bm{\partial}_{(3)} \Psi_1 + 14 H_{\mathrm{har}} \mathbf{N}_{\mathrm{har,(3)}} + 2 \dot{\mathbf{N}}_{\mathrm{har,(3)}}\right) + 6 \dot{H}_{\mathrm{har}} \mathbf{N}_{\mathrm{har,(3)}} \\
    \qquad = 12 \pi G_{\mathrm{N}} a^6(\bar{\rho} + \bar{p}) \bm{v}_{\mathrm{har,(3)}}, \\[8pt]
{-3\Box_{\mathrm{har}} \left(a^6 \mathcal{V}_{\mathrm{har,(3)}}\right) + {84} H_{\mathrm{har}}^2 \mathcal{V}_{\mathrm{har,(3)}} + 12 H_{\mathrm{har}} \left(2 \mathcal{N}_{(3)} + \dot{\mathcal{V}}_{\mathrm{har,(3)}}\right) + 6 \dot{H}_{\mathrm{har}} \mathcal{V}_{\mathrm{har,(3)}} = \mathbf{0}_{(3\times 3)},}
\end{cases}
\end{equation}
where $\Box_{\mathrm{har}}$ denotes the Laplace--Beltrami operator corresponding to $\mathcal{G}_{\mathrm{FLRW}}^{\mathrm{har}}$, $\bm{\partial}_{(3)} \coloneqq (\partial_x, \partial_y, \partial_z)^T$ is the spatial gradient vector, and dots represent derivatives with respect to harmonic time $t$. Here, $\bm{v}_{\mathrm{har,(3)}} \coloneqq \frac{1}{a^6}\frac{d\mathbf{r}^T}{dt}$ is the spatial harmonic three-velocity, and $\mathcal{N}_{(3)}$ denotes the \emph{pure shear matrix}, defined as the off-diagonal symmetric part of $\bm{\partial}_{(3)}\mathbf{N}_{\mathrm{har,(3)}}^T$:
\begin{equation}
    \mathcal{N}_{(3)} \coloneqq \operatorname{Sym}\left(\bm{\partial}_{(3)} \mathbf{N}_{\mathrm{har,(3)}}^T\right) - \bm{\partial}_{(3)} \mathbf{N}_{\mathrm{har,(3)}}^T \odot \operatorname{Id}_{(3)}.
\end{equation}


{
\begin{remark}
In the static background limit---characterized by $a=1$ (and hence ${H}_{\mathrm{har}}={0}$), $\bar{\rho}=0$, and $\bar{p}=0$---the dynamical equations reduce to the wave system:
\begin{equation}\label{eq:harmonic_perturbation_system}
\begin{cases}
    \Box \Psi_1 = -4\pi G_{\mathrm{N}} (\delta \rho + 3\delta p), \\[6pt]
    \Box \Psi_2 = -4 \pi G_{\mathrm{N}} (\delta\rho - \delta p), \\[6pt]
    \Box \mathbf{N}_{\mathrm{har},(3)} = \mathbf{0}_{(3)}, \\[6pt]
    \Box \mathcal{V}_{\mathrm{har},(3)} = \mathbf{0}_{(3\times 3)},
\end{cases}
\end{equation}
which precisely match Eqs.~\eqref{eq_lin} upon setting $\rho = \delta\rho$ and $p = \delta p$, treating $\bm{v}_{(3)}$ as a first-order perturbation, retaining strictly linear terms, and identifying the scalar  potentials via $\Psi_1 \equiv \Phi_1$ and $\Psi_2 \equiv \Phi_2$.
\end{remark}

}

Furthermore, requiring the coordinate functions to satisfy the harmonic gauge condition at linear order yields the following constraint relations connecting the spatial derivatives of the vector and matrix potentials to the scalar potentials:
{
\begin{equation}
\begin{cases}
    4 a^4 \bm{\partial}_{(3)}^T \mathbf{N}_{\mathrm{har,(3)}} = -12 H_{\mathrm{har}} (\Psi_1 + \Psi_2) - \dot{\Psi}_1 - 3 \dot{\Psi}_2, \\[8pt]
    4 a^4 \bm{\partial}_{(3)}^T \mathcal{V}_{\mathrm{har,(3)}}+\frac{1}{a^2}\bm{\partial}^T_{(3)}(\mathcal{H}_{(3)}\odot \mbox{Id}_{(3)}) = -32 H_{\mathrm{har}} \mathbf{N}_{\mathrm{har,(3)}}^T - 4 \dot{\mathbf{N}}_{\mathrm{har,(3)}}^T + \bm{\partial}_{(3)}^T(3 \Psi_2 - \Psi_1).
\end{cases}
\end{equation}}

Finally, the equations of motion for the perturbed matter source—corresponding to energy and momentum conservation $\operatorname{div}(\mathfrak{T}) = 0$—are expressed in compact matrix divergence form as
\begin{equation}
2\bm{\partial}^T\left(\sqrt{-g}\mathcal{G}^\sharp \mathcal{T}\right) - \sqrt{-g}\left(\operatorname{tr}\left(\mathcal{T}^\sharp \bm{\partial}\mathcal{G}\right)\right)^T = \bm{0}^T.
\end{equation}

}

{

\section{Conclusions}\label{sec:conclusions}

In this work, we have developed a self-contained, index-free coordinate-matrix framework for General Relativity formulated in the harmonic gauge. By mapping second-order covariant and contravariant physical tensors onto the space of $4\times 4$ real matrices $M_4$, we have established a concise algebraic formalism underpinned by musical isomorphisms ($\sharp, \flat$), metric adjoints ($\dagger$), and matrix-valued derivative vectors ($\bm{\mathcal{B}}, \bm{\mathcal{C}}, \bm{\mathcal{D}}^\sharp$).

A central feature of this formalism is the transparent matrix representation of the Laplace--Beltrami operator in harmonic coordinates as a matrix trace, $\Box_{\mathrm{LB}} \coloneqq \operatorname{tr}(\mathcal{G}^\sharp \partial^2)$, where $\partial^2 \coloneqq \bm{\partial}\bm{\partial}^T$ represents the matrix Hessian operator of second partial derivatives. Within this setting, the non-linear structure of the Ricci matrix $\mathcal{R}$ decomposes naturally into symmetric trace combinations of self-adjoint and anti-self-adjoint matrix vectors ($\bm{\mathcal{C}}^{(+)}, \bm{\mathcal{C}}^{(-)}$), highlighting the intrinsic linear-algebraic symmetries of curved spacetimes.

Furthermore, we have demonstrated that Einstein's field equations assume the direct form of a non-linear matrix wave equation for the metric matrix $\mathcal{G}$, namely $\Box_{\mathrm{LB}}\mathcal{G} = -16\pi G_{\mathrm{N}} \hat{\mathcal{P}}_-(\mathcal{T} + \bar{\mathcal{T}})$. This representation isolates the gravitational energy-momentum pseudotensor matrix $\bar{\mathcal{T}}$ and enables an explicit, index-free derivation of its exact algebraic relationship with the classical Landau--Lifshitz energy-momentum complex $\mathcal{T}_{\mathrm{LL}}$. Consequently, the total local energy-momentum conservation law is expressed in the compact matrix divergence form $\bm{\partial}^T \left((-g)(\mathcal{T}^{\sharp} + \mathcal{T}_{\mathrm{LL}}^{\sharp})\right) = \bm{0}^T$.

The analytical power and versatility of our matrix framework have been validated through three key physical applications:
\begin{enumerate}
    \item \emph{Linearized gravity:} The matrix wave equation straightforwardly reproduces the linear wave dynamics of metric perturbations in weak fields.
    \item \emph{FLRW cosmology:} The full matrix field equations yield the exact Friedmann cosmological dynamics parametrized by harmonic time.
    \item \emph{Cosmological perturbation theory:} Matrix-adapted gradient and divergence operators provide a unified, index-free formulation of linear scalar perturbation equations and fluid conservation laws in both the Newtonian and harmonic gauges.
\end{enumerate}

Beyond its formal compactness, the proposed index-free matrix framework may provide a useful basis for symbolic implementations and further developments in gravitational physics. Potential applications include numerical relativity, gravitational-wave propagation, higher-order post-Newtonian expansions, and extensions to alternative theories of gravity.


}

\begin{acknowledgments}
This work  is supported by the Spanish grant PID2025-168310NB-I00 funded by MICIU/AEI/10.13039/501100011033 and by FSE+.
\end{acknowledgments}

\end{document}